\documentclass[aip,graphicx]{revtex4-1}
\usepackage{graphicx}
\newcommand{\angstrom}{\mbox{\normalfont\AA}}
\draft 

\begin{document}


\title{Controlling Projection‑Space Artifacts in DFT+U via Projection‑Consistent $U_{\mathrm{eff}}$} 



\author{Manjula Raman}

\author{Kenneth Park}%
 \email{kenneth\_park@baylor.edu.}
 \homepage{https://physics.artsandsciences.baylor.edu/person/dr-kenneth-t-park}
\affiliation{ 
Department of Physics \& Astronomy, Baylor University, Waco, Texas 76798-7316, USA 
}%



\date{\today}

\begin{abstract}
Density functional theory augmented with a Hubbard correction (DFT+U) is widely used to treat localized electronic states, but its predictions are often sensitive to the choice of the local projection space defining the correlated subspace. This sensitivity poses a practical challenge for computational reproducibility, particularly when projection parameters vary across codes, basis sets, or materials. In this work, we systematically investigate how the effective on-site Coulomb interaction $U_{\mathrm{eff}}$, determined \textit{ab initio} using constrained density functional theory, depends on the size of the local projection space in all-electron APW+lo calculations. Using rutile and anatase TiO$_2$ and $\beta$-MnO$_2$ as representative test cases, we show that applying a single fixed $U_{\mathrm{eff}}$ across different projection choices introduces artificial projection-driven errors in total energies, including spurious magnetic ordering transitions and unphysical sensitivity of phase stability. These artifacts are eliminated when $U_{\mathrm{eff}}$ is determined in an internally consistent manner for each projection space, yielding projection-consistent DFT+U predictions for lattice parameters, phase energetics, and magnetic ground states. By analyzing total-energy trends alongside the spatial characteristics of the localized $d$ orbitals, we demonstrate that the systematic reduction of $U_{\mathrm{eff}}$ with increasing projection size originates from orbital relaxation and enhanced electronic screening associated with orbital spatial extension. These results provide a physically motivated framework for controlling projection-space artifacts in DFT+U calculations and for obtaining energetically robust predictions across diverse correlated materials and computational setups.
\end{abstract}


\maketitle 


\section{\label{sec:level1}Introduction}
The density functional theory supplemented with Hubbard-correction (DFT+U) is widely used to study systems with strong electron correlation effects \cite{Anisimov1997,himmetoglu2014hubbard}. Its popularity stems from the substantial improvements it offers---particularly in predicting electronic structure---while maintaining relatively modest computational cost. However, the accuracy of DFT+U results is highly sensitive to several user-defined input parameters and methodological choices. Among these, the on-site Coulomb energy $U$ and exchange energy $J$ play central roles, as they directly influence the effective electron–electron interactions in solids. This sensitivity has led to widespread practice of tuning $U$ to reproduce experimental observation of selected properties. 

Another important consideration is whether to use separate $U$ and $J$ values or the combined parameter $U_{\mathrm{eff}}$ $=U-J$ (with $J =$ $0$).
This choice can significantly affect outcomes. For example, Tompsett \textit{et al.} \cite{Tompsett2012} showed that the “full anisotropy” approach--explicitly including both $U$ and $J$--correctly predicts the antiferromagnetic ground state and the $0.8$ eV band gap of $\beta$-MnO$_2$. In contrast, the simplified $U_{\mathrm{eff}}$ approach incorrectly favors a ferromagnetic state. They attributed this improvement to a more accurate description of spin polarization in Mn $d$ orbitals, particularly through reducing unphysical spin polarization of bonding states. Similarly, Mellan \textit{et al.} \cite{Mellan2015} demonstrated that the correct magnetic ground state and Jahn–Teller distortion in LaMnO$_3$ are reproduced only  when the exchange term $J$ is explicitly included. Yoon \textit{et al.} \cite{Yoon2021} further reported that the exchange correction has a stronger influence in Mn$^{3+}$ than in Mn$^{2+}$ in their study of mixed-valent Mn$_3$O$_4$.  

Beyond the choice of $U$ and $J$, many studies have noted that the size of the projection sphere and the construction of local orbitals can strongly influence quantitative results—from lattice parameters to electronic structures as well as magnetic ground states  \cite{wang2016local,wang2017implications,pickett1998reformulation,fabris2005electronic,tablero2008representations,geneste2017dft+,outerovitch2023electronic,park2010electronic,park2015density,Park2024}. Wang \textit{et al.} \cite{wang2016local} reported substantial discrepancies in DFT+U calculations of MnO and MnO$_2$ using the $U_{\mathrm{eff}}$ scheme across different computational codes. They attributed these differences to variations in the radial extent of Mn $3d$ orbitals defining the Hubbard subspace. In our recent work \cite{Park2024}, we systematically demonstrated that the spatial range used for the Hubbard correction alone can lead to strikingly different results in TiO$_2$. Notably, the optimized lattice parameters exhibit strong sensitivity to the input U, with the lattice constant increasing by as much as $1.8$ \%. In addition, when the projection space for the Hubbard correction is chosen too small, even large $U$ values (up to $10$ eV) fail to stabilize rutile against anatase, contrary to established literature \cite{dompablo2011dft+, curnan2015investigating}. These discrepancies arise from the dependence of orbital occupancy on the projection size. The differing orbital occupancy further affects the orbital-dependent potential energy shifts and produces quantitatively distinct electronic structures. Careful consideration is therefore essential for interpreting DFT+U results and extracting meaningful physical insights.

The sensitivity of DFT+U to the size of the local projection space is problematic because this space is not uniquely defined, yet orbital occupations---and thus the Hubbard correction---depend on it. To address this issue, Wang and Jiang \cite{wang2019local} reformulated DFT+U using a Thomas--Fermi screening model to determine $U$, reporting reduced sensitivity to projection size in several oxides. Nawa \textit{et al.} \cite{nawa2018scaled} made an important advance by using linear response theory \cite{cococcioni2005linear,pickett1998reformulation} to show that $U_{\mathrm{eff}}$ decreases by $2$--$3$ eV for transition-metal monoxides as the muffin-tin radius is increased. Despite the significant changes in $U_{\mathrm{eff}}$, they obtained valence-band structures that varied little employing internally consistent $U_{\mathrm{eff}}$ values. They attributed this "invariance" to a simple scaling argument, the approximately linear relationship between $U_{\mathrm{eff}}$ and the occupancy of local orbitals.  

The work of Nawa \textit{et al.} establishes that $U_{\mathrm{eff}}$ values may not be freely transferred across different computational setups, and that consistency in calculated results can be achieved using appropriately scaled values. However, important questions remain. First, does a similar linear dependence on projection size appear when $U_{\mathrm{eff}}$ is estimated via an alternative \textit{ab initio} method? In contrast to the linear response theory, the constrained DFT method proposed by Anisimov and co-workers \cite{AnisimovGunnarsson1991} prescribes specific local orbital occupations to estimate the Coulomb interaction and decouples the localized orbitals from the rest of the basis set. While the two different methods are believed to be complementary, confirming a similar projection-size dependence in this framework would broaden its applicability across computational approaches. Second, it was not clear without further physical insight whether the scaling relationship extends beyond transition-metal monoxides to strongly correlated systems with more pronounced covalent bonding characters. Finally, whether consistent results can be achieved in other sensitive properties—such as optimized lattice parameters and relative phase energetics—remains an open and practically important question, given how strongly these properties depend on the choice of $U_{\mathrm{eff}}$. 

In this paper, we systematically investigate how $U_{\mathrm{eff}}$ varies with projection size, the physical mechanism underlying this variation, and its consequences for DFT+U predictions. We use the constrained DFT method to determine $U_{\mathrm{eff}}$, examining whether the linear dependence on projection size observed via linear response theory also holds in this distinct framework. We further demonstrate that using a fixed $U_{\mathrm{eff}}$ calibrated at one projection size introduces artificial, projection-size-driven errors in calculated properties, and that these artifacts vanish when $U_{\mathrm{eff}}$ is recalculated for each projection space. We refer to this as the projection-consistent $U_{\mathrm{eff}}$ scheme, in which $U_{\mathrm{eff}}$ is determined via constrained DFT for each chosen projection space, ensuring internal consistency between the Hubbard correction and its projection space. We illustrate these effects using rutile and anatase TiO$_2$ and $\beta$-MnO$_2$—materials of broad technological relevance in heterogeneous catalysis, photocatalysis, and energy storage, whose structural, electronic, and energetic properties are known to be sensitive to the Hubbard projection space \cite{chen2011increasing,matsubu2017adsorbate,guo2019fundamentals,lang2020single,jiao2007mesoporous,debart2008alpha}. Two particularly demanding and practically important predictions are tested: the relative phase stability between rutile and anatase TiO$_2$ and the magnetic ground-state ordering in $\beta$-MnO$_2$. Finally we show that the decrease of $U_{\mathrm{eff}}$ with increasing projection size arises from orbital relaxation and enhanced screening as the spatial extent of the localized d orbitals expands—a mechanism consistent with the analysis of Solovyev and Dederichs \cite{Solovyev1994} for transition-metal impurities.

\section{Methods and Computational Details}
\subsection{\label{sec:level2}LAPW and APW+lo approaches}

The DFT calculations were performed using the all-electron, full-potential augmented plane wave plus local orbitals (APW+lo) method implemented in the WIEN2k software package \cite{Blaha2001,Singh2006}. In this method, the unit cell is divided into non-overlapping muffin-tin (MT) spheres centered at atomic sites and an interstitial region. Inside the MT spheres, the Kohn-Sham wave functions are expanded in spherical harmonics multiplied by radial functions, while in the interstitial region they are represented by plane waves.

The linearized APW (LAPW) basis functions within the atomic spheres are expressed as follows: 

\begin{equation}
\psi_{l,m}^{\mathrm{MT}}(r) = A_{l,m} u_{l}(r,E_{l}) + B_{l,m} \dot{u}_{l}(r,E_{l})
\label{eq:LAPW}
\end{equation}

where $u_{l}(r, E_{l})$ is the solution to the radial Schrödinger equation at energy $E_{l}$, and $\dot{u}_{l}(r, E_{l})$ is its energy derivative. The coefficients $A_{l,m}$ and $B_{l,m}$ are determined to ensure that the LAPW basis functions match smoothly with the interstitial plane waves, preserving both the function and its derivative at the sphere boundaries.

\medskip
\medskip

To improve variational flexibility of the basis set---particularly for semi-core and higher-lying conduction states---the LAPW basis was augmented with local orbitals (lo), forming the APW+lo scheme. \cite{Madsen2005} The local orbitals are defined as 

\begin{equation}
\varphi_{l,m}^{\mathrm{lo}}(r) = A_{l,m}^{\mathrm{lo}} u_{l}(r,E_{l}^{\mathrm{lo}}) + B_{l,m}^{\mathrm{lo}} \dot{u}_{l}(r,E_{l}^{\mathrm{lo}})
\label{eq:APWlo}
\end{equation}

and are normalized within the sphere, while vanishing at the MT boundary. These orbitals enhance the description of states not well represented by the standard LAPW basis. 

All calculations employed the APW+lo scheme. Core and valence electrons were separated at an energy threshold of –6.0 Ry. For titanium, the 3$s$ and 3$p$ were treated as semi-core, while the Ti 3$d$, 4$s$, and O 2$s$, 2$p$ were treated as valence. Core states were handled fully relativistically via the Dirac–Fock approach, and valence states were treated with a scalar-relativistic approximation. Exchange-correlation effects were described by the Perdew–Burke–Ernzerhof (PBE) functional\cite{Perdew1996,Perdew1998} within the generalized gradient approximation (GGA).

The basis-set size of our calculations was controlled through the dimensionless parameter \( R_{\mathrm{MT}} \)\( K_{\mathrm{Max}} \). To balance accuracy and computational cost, we initially used the Wien2k  "precision level 2," corresponding to \( R_{\mathrm{MT}} \)\( K_{\mathrm{Max}} \) values from $7.00$ to $7.22$ depending on the chosen MT radii. These values, along with the number of $k$-points in Brillouin zone and the cutoff $g_{max}$ for the Fourier expansion of the charge density and potential, were subsequently increased during convergence testing. Self-consistency was reached when the total energy converged within $0.000001$ Ry and charges within $0.00001$. For structural relaxations, atomic forces were converged to below $1$ mRy/$a_B$. 

\subsection{\label{sec:level2} DFT+U Calculations and Control of Projection Space}
The Hubbard-corrected DFT+$U$ method is implemented within the rotationally invariant formalism, in which an explicit on-site interaction term is added for a selected set of localized orbitals defined inside the muffin-tin spheres.\cite{anisimov1993density} The correlated subspace is constructed from atomic-like orbitals, and the occupation matrices are obtained by projecting the Kohn--Sham states onto these localized functions.\cite{Madsen2005} In the commonly used simplified approach, the on-site Coulomb and exchange interactions are combined into a single effective parameter $U_{\mathrm{eff}} = U - J$, leading to the Dudarev energy correction
\begin{equation}
E_U
=
\frac{U_{\mathrm{eff}}}{2}
\sum_{I,\sigma}
\mathrm{Tr}
\left[
\mathbf{n}^{I\sigma}
\left(
\mathbf{1}
-
\mathbf{n}^{I\sigma}
\right)
\right].
\label{eq:Hubbard}
\end{equation}

Here $\mathbf{n}^{I\sigma}$ is the occupation matrix of the correlated localized orbitals on site $I$ for spin $\sigma$, obtained by projecting the Kohn--Sham states onto the chosen Hubbard subspace within the muffin-tin sphere. The trace is taken over the orbital indices spanning this subspace.\cite{dudarev1998electron} In this rotationally invariant formulation, the interaction and double-counting contributions are combined into the single correction term given above. This orbital-dependent correction penalizes fractional occupations and energetically favors localized integer filling.\cite{novak2006exact} Within the LAPW/APW+lo framework, the projection of the Kohn--Sham states is confined to the muffin-tin spheres, so the definition of the correlated subspace---and consequently the orbital occupancies and Hubbard energy---depends explicitly on the choice of muffin-tin radii.

To examine projection-size effects, we systematically varied the MT radius \( R_{\mathrm{MT}} \). For Ti in rutile and anatase, \( R_{\mathrm{MT}} \) was increased from $1.91$ to $2.00$, $2.10$, $2.20$, and $2.30$ bohr radius \(a_B \). To prevent overlap of MT spheres, the oxygen 
radii were reduced from $1.59$ to $1.50$, $1.40$, and $1.30$ \(a_B \); the latter value was used for Ti $R_{\mathrm{MT}}$ $=$ $2.20$, and $2.30$ $a_B$. A similar procedure was applied to $\beta$-MnO$_2$, where Mn \( R_{\mathrm{MT}} \) was varied from $1.90$ to $2.00$, $2.10$, and $2.20$ \(a_B \) and O \( R_{\mathrm{MT}} \) was adjusted from $1.55$ to $1.48$, $1.40$, and $1.30$ \(a_B \). 

The magnetic ground state of $\beta$-MnO$_2$ is known to be a helical antiferromagnetic (AFM) with pitch ($7/2$) along the $c$ axis.\cite{yoshimori1959new,ohama1971determination,regulski2003incommensurate} For the comparison in the total energy and other properties between AFM and ferromagnetic (FM) configurations, we adopted a simplified collinear antiferromagnetic ordering---the magnetic moments aligned in opposite directions along the body diagonal. This approach has been commonly used in the literature to assess Hubbard-corrected energetics and property calculations in MnO$_2$ \cite{franchini2007ground,Tompsett2012,wang2016local, mahajan2021importance}.  

\subsection{\label{sec:level2}Estimating $U_{\mathrm{eff}}$ \textit{ab initio} with constrained DFT}
The on-site Coulomb energy was calculated using the constrained DFT (cDFT) method proposed by Anisimov and Gunnarsson. \cite{Anisimov1991,AnisimovGunnarsson1991} In this approach, $U_{\mathrm{eff}}$ is interpreted as the energy cost of transferring a  localized $d$-- or $f$-- electron from one site to another: $d^n + d^n \rightarrow d^{n+1} + d^{n-1}$. The effective Hubbard parameter is approximated as the lowest-order screened Slater integral $F^0$:

\begin{equation}
U_{\mathrm{eff}} = \varepsilon_{3d^\uparrow}\left(\frac{n+1}{2}, \frac{n}{2}\right) - \varepsilon_{3d^\uparrow}\left(\frac{n+1}{2}, \frac{n}{2}-1\right) - \varepsilon_F\left(\frac{n+1}{2}, \frac{n}{2}\right) + \varepsilon_F\left(\frac{n+1}{2}, \frac{n}{2}-1\right)
\label{eq:Ueff}
\end{equation}
where \(\varepsilon_{3d^\uparrow}(n^\uparrow, n^\downarrow)\) is the spin-up 3$d$ eigenvalue for the specified occupancy, and \(\varepsilon_F(n^\uparrow, n^\downarrow)\) is the corresponding Fermi energy. 

In the original formulation using a linearized MT orbital basis, the hopping integrals between the impurity $d$ states and the rest of the system were set to zero. Because this constraint cannot be applied directly within the LAPW/APW+lo framework, we followed the method proposed by Madsen and Novak \cite{Madsen2005}, in which the linearization energy is placed far above the Fermi level to effectively remove the $d$ orbitals from the basis. For TiO$_2$, $2\times2\times2$ and $2\times2\times3$ supercells were constructed using fully optimized lattice parameters, with one site designated as the impurity. For $\beta$-MnO$_2$, experimental lattice parameters were used to build a $2\times2\times2$ supercell so that our calculated $U_{\mathrm{eff}}$ values could be compared to previously reported results. \cite{Tompsett2012} Detailed procedures and convergence tests are provided in Table S1 of the Supplementary Materials. 

\section{Results}
\subsection{Properties of TiO$_2$ Calculated using Projection-consistent $U_{\mathrm{eff}}$ Scheme}
\begin{figure}[h]
    \centering
    \includegraphics[width=0.8\textwidth]{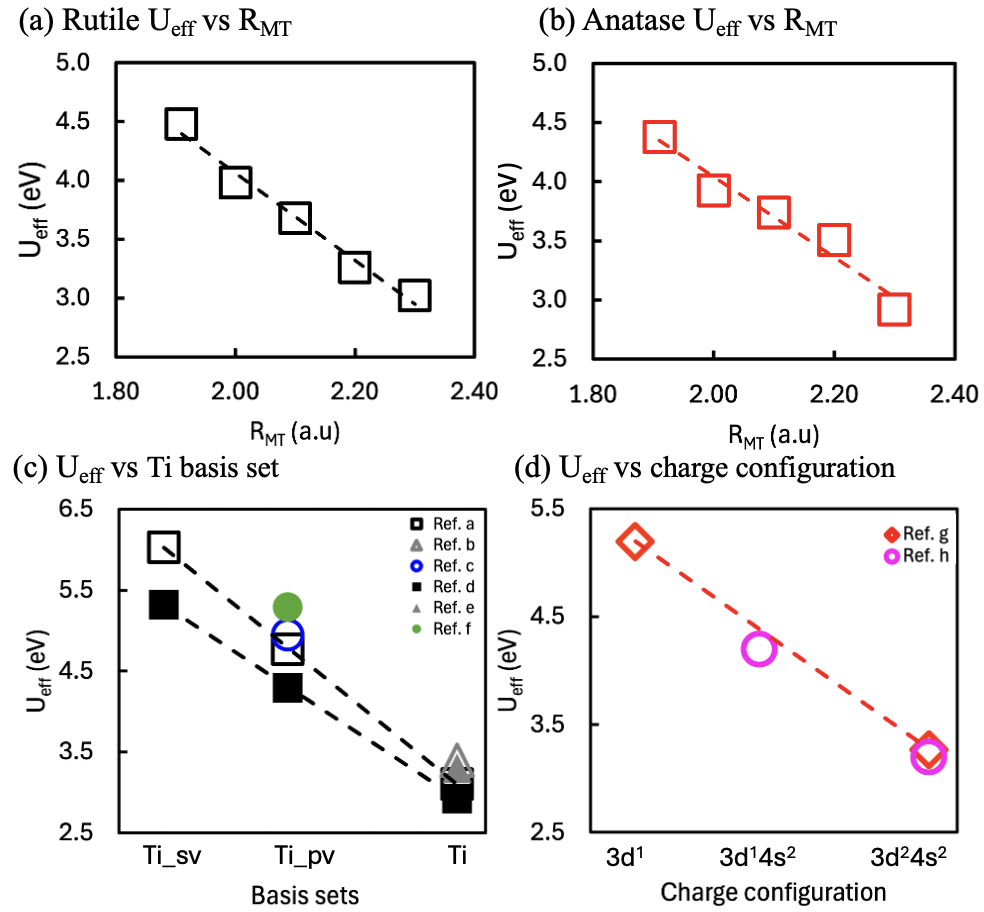}
       \caption{(a) The self-consistently determined on-site Coulomb interaction parameter \(U_{\mathrm{eff}}\) as a function of Ti muffin-tin radius \(R_{\mathrm{MT}}\) for rutile TiO$_2$ using the cDFT method. (b) Same as (a), but for anatase TiO$_2$. \(U_{\mathrm{eff}}\) values obtained using linear-response theory for rutile (open symbols) and anatase (filled markers) in literature using (c) different Ti valence sets and (d) Ti charge configuration. Ti, Ti\_pv, and Ti\_sv represent standard, \textit{p}-- and \textit{s}--semicore inclusive pseudopotentials, respectively. 
    The dashed lines are only to guide the eyes. references a and d: \cite{curnan2015investigating}, b and e: \cite{Mattioli2008}, c: \cite{xu2015linear}, f: \cite{Raghava2022}, g: \cite{Orhan2020}, h: \cite{farnesi2011ideal}.}
    \label{fig:UvsRmt}
\end{figure}

Figure \ref{fig:UvsRmt} (a) shows the projection-consistent $U_{\mathrm{eff}}$ values for rutile TiO$_2$ computed at several Ti muffin-tin radii. The value is largest, $4.5$ eV at $R_{MT}=$ $1.91$ $a_B$. As $R_{MT}$ increases, $U_{\mathrm{eff}}$ decreases steadily, reaching $3.0$ eV at $R_{MT}=$ $2.3$ $a_B$, roughly $67$ $\%$ of the initial value. A similar trend is observed for anatase, where $U_{\mathrm{eff}}$ decreases from $4.4$ eV at $R_{\mathrm{MT}}$ = $1.91$ $a_B$ to $3.0$ eV at $2.3$ $a_B$ (Fig. \ref{fig:UvsRmt}(b)). This clear monotonic reduction demonstrates an unmistakable dependence of the effective Coulomb interaction on the muffin-tin radius, just as Nawa and co-workers reported in their study of transition metal monoxides. \cite{nawa2018scaled}

To determine whether a similar trend appears in other \textit{ab initio} approaches, several literature values obtained via the linear-response theory\cite{cococcioni2005linear,pickett1998reformulation} using pseudopotential methods are compiled in Figs. \ref{fig:UvsRmt} (c) \& (d). Two observations emerge. First, $U_{\mathrm{eff}}$ tends to be larger when the pseudopotentials include semi-core states into the valence states. For instance, Curnan and Kitchin \cite{curnan2015investigating} reported a value of $3.102$ eV for rutile using standard PAW pseudopotentials with PBE (open squares). Including $p$ semi-core states increased this value to $4.773$ eV, and  including $s$ semi-core states raised it further to $6.030$ eV. They noted that the standard PAW-pseudopotentials result agrees well with the value reported by Mattioli and Filippone \textit{et al.}---$3.23$ eV--- obtained with standard ultrasoft pseudopotentials (grey open triangle). \cite{Mattioli2008,mattioli2010deep} On the other hand, the $U_{\mathrm{eff}}$ value obtained with $p$-inclusive PAW pseudopotentials closely matched the $4.95$ eV reported by Xu \textit{et al.} using a pseudopotential constructed from GBRV library (blue open circle) \cite{xu2015linear}. They further observed a similar trend of semi-core inclusive pseudopotentials producing larger $U_{\mathrm{eff}}$ for anatase as well (filled squares). Additional data collected in Fig. \ref{fig:UvsRmt} (c) support this observation (filled triangle and circle).

Second, $U_{\mathrm{eff}}$ also depends strongly on the ionic configuration used to generate the pseudopotential. Orhan and O’Regan \textit{et al.}\cite{Orhan2020} employed a minimum-tracking variant of linear response scheme with LDA functional to evaluate $U_{\mathrm{eff}}$.
They reported values of $\sim 3.27$ eV for rutile and anatase when using the pseudopotential constructed from a neutral Ti configuration (red diamonds in Fig. \ref{fig:UvsRmt} (d)). However, switching to a Ti$^{3+}$ reference state ($3d^1$) increased $U_{\mathrm{eff}}$ to $5.20$ eV. They attributed the larger $U_{\mathrm{eff}}$ to the more localized Ti $3d$ subspace represented by this pseudopotential. Likewise, Camellone \textit{et al.} \cite{farnesi2011ideal} obtained $U_{\mathrm{eff}} =$ $3.2$ eV for rutile TiO$_2$ using a PAW-based pseudopotential generated from a neutral atom configuration with PBE (pink circles). When using a pseudopotential constructed from a Ti$^{1+}$ reference state ($3d^1 4s^2$), their calculated $U_{\mathrm{eff}}$ value increased to $4.2$ eV. They also noted a similar sensitivity in FeO, where $U_{\mathrm{eff}}$ increases from $4.6$ eV for a pseudopotential created from a neutral Fe atom, to $7.8$ eV for one generated from an Fe$^{2+}$ reference state in the LDA+U calculations employing a linear combination of atomic orbital basis\cite{pickett1998reformulation}.  

Although the absolute magnitudes of $U_{\mathrm{eff}}$ vary among the linear-response and the constrained-occupancy approaches, the observed dependencies on pseudopotential valence configuration and charge state closely mirror the $R_{MT}$ dependence Nawa \textit{et al.} \cite{nawa2018scaled} reported as well as we observed (Figs. \ref{fig:UvsRmt} (a) \& (b)). The consistency across different computational frameworks--whether using LAPW/APW+lo or pseudopotential methods, and whether applying cDFT or linear response theory--suggests that the results unlikely originate from any specific methodology. Instead, it points to a common underlying physical cause that all approaches are capturing. Since the size of Hubbard projection in the LAPW/APW+lo approach is independently controlled by a single parameter--the muffin-tin radius, it is reasonable to conclude that the sensitivity of $U_{\mathrm{eff}}$ on the differences in pseudopotential construction also stems from the spatial extent of the localized $d$ subspace.

\begin{figure}[h!]
    \centering
    \includegraphics[width=0.8\textwidth, keepaspectratio]{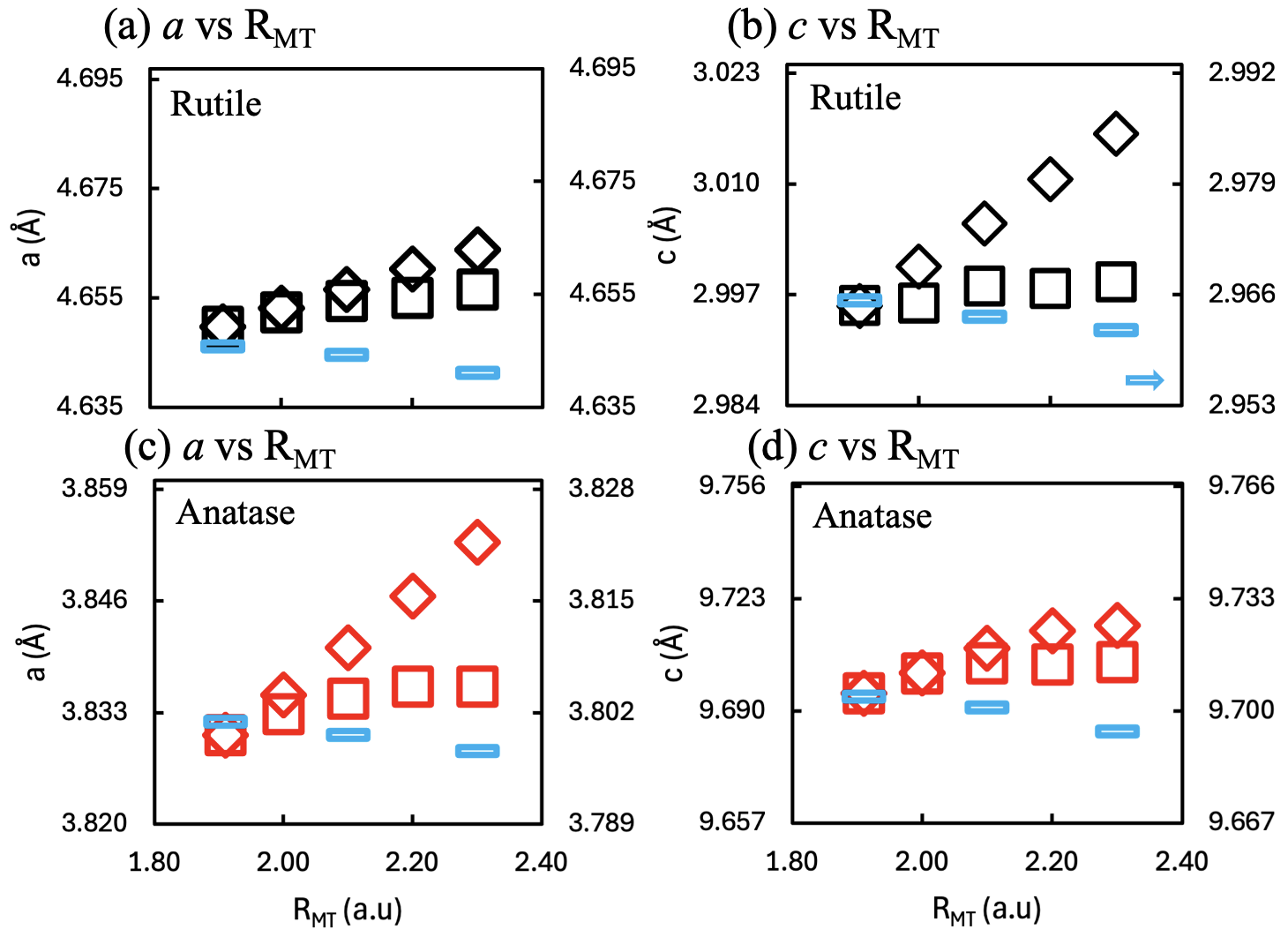}
    \caption{Calculated lattice parameters (a) $a$ and (b) $c$ as a function of Ti muffin-tin radius $R_{\mathrm{MT}}$ for rutile and (c) \& (d) for anatase TiO\(_2\), with two computational schemes: fixed $U_{\mathrm{eff}}$ (diamond markers) and projection-consistent $U_{\mathrm{eff}}$ (square markers) schemes. For actual $U_{\mathrm{eff}}$ values used, see the main text and Table S2 in the Supporting Materials. As for the reference values, the lattice parameters computed without the Hubbard correction (light blue) are also shown for comparison.}  
    \label{fig:latticevsRmt}
\end{figure} 

For several calculated properties of TiO$_2$ by the DFT+U method, we compare two schemes for applying the on-site interaction parameter $U_{\mathrm{eff}}$. Under the fixed $U_{\mathrm{eff}}$ scheme, $U_{\mathrm{eff}}$ values are fixed at $4.4$ and $4.3$ eV for rutile and anatase, respectively, regardless of the chosen $R_{\mathrm{MT}}$ values. These values correspond to the $R_{\mathrm{MT}} = 1.91~a_{\mathrm{B}}$ intercepts obtained from the linear fits shown in Fig. \ref{fig:UvsRmt}, and were used to eliminate residual numerical variations arising from supercell size and local-environment differences between the two polymorphs. In contrast, in the projection-consistent $U_{\mathrm{eff}}$ scheme, $R_{\mathrm{MT}}$-dependent $U_{\mathrm{eff}}$ values obtained from the same linear fits were used for each projection space (Table S2), rather than applying a single fixed value. This approach follows the same projection-scaling philosophy introduced by Nawa \textit{et al.}, but is implemented here within the constrained DFT framework and explicitly enforces internal consistency between the Hubbard correction and the chosen projection space.

Figure \ref{fig:latticevsRmt} first presents the optimized lattice parameters as a function of Ti muffin-tin radius $R_{\mathrm{MT}}$ for rutile and anatase TiO$_2$, calculated using the two $U_{\mathrm{eff}}$ schemes described above. Under the fixed-U scheme (diamonds), the lattice parameter $c$ expands by about $0.7$ \% as $R_{\mathrm{MT}}$ increases from $1.91$ to $2.3$ $a_B$ while the lattice parameter $a$ shows a smaller increase of approximately $0.3$ \%. This asymmetrical lattice expansion has been noted previously.\cite{Park2024} It has been attributed to the uneven bond lengths in rutile structure combined with increased Ti ionic character especially at larger $R_{\mathrm{MT}}$, arising from greater depletion of the $d$-orbital population ($n_d$) (Table S3). In contrast, structural optimization using the projection-consistent $U_{\mathrm{eff}}$ values for each projection space yields lattice parameters that remain largely consistent across different $R_{\mathrm{MT}}$ values (squares). Although the resulting lattice constants are slightly larger than those obtained using PBE alone (blue bars), as expected, they vary no more than $0.006$ $\angstrom$ across the full range of $R_{\mathrm{MT}}$. Their consistency is comparable to the PBE reference without the Hubbard correction. 

Despite the substantial improvement afforded by the projection-consistent $U_{\mathrm{eff}}$ scheme, the lattice parameters especially calculated at larger $R_{\mathrm{MT}}$ values exhibit some residual sensitivity to the projection size. The exact origin of this deviation is unclear. One possibility is that, as the difference between the Ti and O in muffin-tin radii grows, the quality of the local basis set expansion gradually deteriorates, reducing consistency with the rest of the data.

\begin{figure}[h]
    \centering
    \includegraphics[width=1.0\textwidth, keepaspectratio]{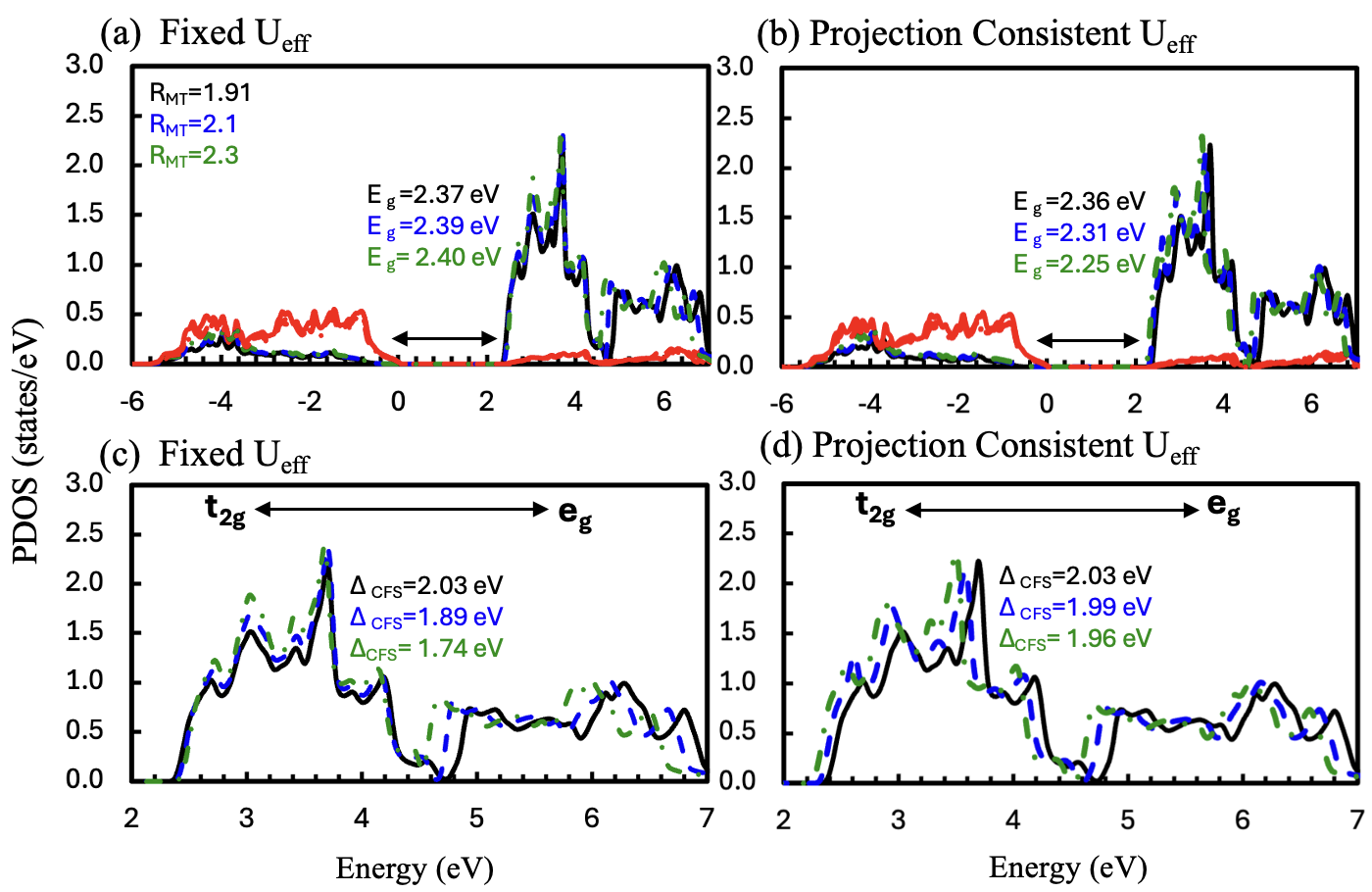}
    \caption{Projected density of states (PDOS) for Ti $d$ and O $p$ orbitals in rutile TiO\(_2\) using (a) fixed and (b) projection-consistent \( U_{\mathrm{eff}} \) schemes for different Ti $R_{\mathrm{MT}}$ : 1.91 (solid black), 2.1 (dashed blue), and 2.3 (dash-dotted green) $a_B$. The PDOS for O 2p is in red with the same line patterns. (c) and (d) panels contrast the difference in the calculated CFS between under the two different schemes.}
    \label{fig:PDOS_rutile_U_cal}
\end{figure}

Next, we examined the electronic structure obtained using the two $U_{\mathrm{eff}}$ schemes. Figure \ref{fig:PDOS_rutile_U_cal} (a) and (b) show the projected density of states (PDOS) for rutile TiO$_2$ in the valence-- and conduction--band regions, computed with fixed and projection-consistent $U_{\mathrm{eff}}$ schemes, respectively. Under the fixed $U_{\mathrm{eff}}$ scheme, the band gap remains nearly constant, increasing only slightly from $2.37$ to $2.40$ eV as Ti $R_{\mathrm{MT}}$ increases. With the projection-consistent $U_{\mathrm{eff}}$, the band gap decreases modestly from $2.36$ to $2.25$ eV (left panels). The overall weak dependence on projection-space size can be attributed in part to the O $p$ character dominating the top of the valence band. In addition, the Ti $t_{2g}$ states that form the bottom of the conduction band are nearly empty and therefore exhibit minimal sensitivity to changes in orbital occupancy. We emphasize that the weak but systematic dependence of the TiO$_2$ band gap on projection size is not alleviated—and is in fact slightly enhanced—when using projection‑dependent $U_{\mathrm{eff}}$, consistent with the behavior reported previously by Orhan and O’Regan. \cite{Orhan2020}

In contrast to the band gap, the crystal‑field splitting (CFS) between the $t_{2g}$ and $e_g$ manifolds shows a markedly different behavior, exhibiting a much stronger and more direct sensitivity to the Hubbard correction and its projection space. Under the fixed $U_{\mathrm{eff}}$ scheme, the CFS decreases by about $0.3$ eV---from $2.03$ to $1.74$ eV---as $R_{\mathrm{MT}}$ increases (Fig. \ref{fig:PDOS_rutile_U_cal}(c)). This reduction arises from uneven shifts of the $t_{2g}$ and $e_g$ states. They are driven by the occupancy dependence of the Hubbard orbital potential, with the bonding $e_g$ orbitals being particularly sensitive on $R_{\mathrm{MT}}$ values \cite{Park2024}. Although the magnitude of this reduction is modest for a $U_{\mathrm{eff}}$ of moderate strength, using a fixed $U_{\mathrm{eff}}$ value that is too large can produce unrealistically small CFS values.\cite{dompablo2011dft+,Park2024,zhao2018does} On the other hand, the projection-consistent $U_{\mathrm{eff}}$ scheme yields a nearly constant CFS, varying only from $2.03$ to $1.96$ eV, owing to the more consistent shifts of the $t_{2g}$ and $e_{g}$ manifolds across different projection sizes (Fig. \ref{fig:PDOS_rutile_U_cal}(d)).  

\begin{figure}[h]
    \centering
    \includegraphics[width=0.75\textwidth, keepaspectratio]{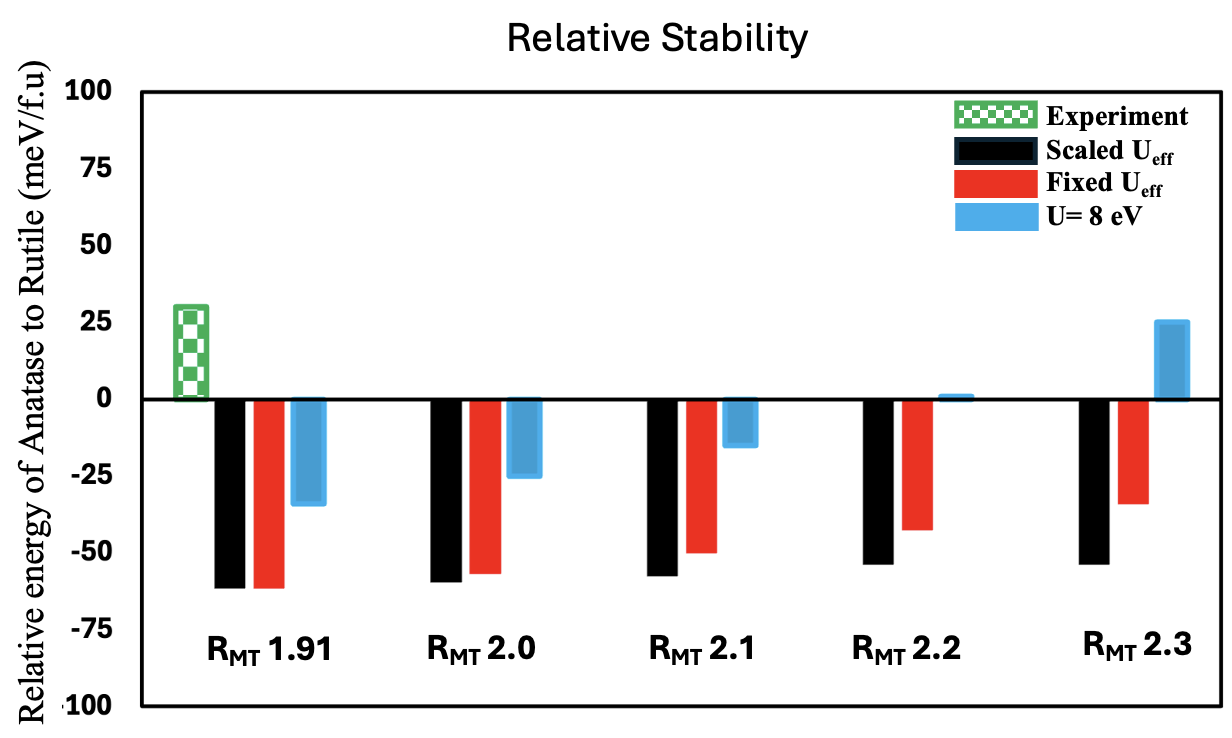}
     \caption{Relative energy difference of anatase with respect to rutile TiO$_2$ as a function of $R_{\mathrm{MT}}$. Calculated using three DFT+U schemes: projection-consistent (scaled) $U_{\mathrm{eff}}$ (black), fixed $U_{\mathrm{eff}}$ (red), and constant $U_{\mathrm{eff}} = 8$ eV (blue). The experimental reference (green) indicates that rutile is the thermodynamic ground state. Negative values indicate anatase is more stable, while positive values indicate rutile is more stable.}
\label{fig:Rutile_Rel_stability}
    \label{fig:Rutile_Rel_stability}
\end{figure}

Another key property of TiO$_2$ that depends strongly on input parameters such as $U_{\mathrm{eff}}$ and $R_{\mathrm{MT}}$ is the relative stability of its two phases. Figure \ref{fig:Rutile_Rel_stability} shows the energy of anatase relative to rutile $(E_R - E_A)$ per formula unit for various $R_{\mathrm{MT}}$ values under different computational schemes. Using the fixed $U_{\mathrm{eff}}$ scheme (red bars), anatase is more stable than rutile by $-62$ meV at $1.91$ $a_B$. However, its stability becomes smaller when a larger $R_{\mathrm{MT}}$ value is used such that anatase is more stable than rutile by $-34$ meV at $2.30$ $a_B$, approximately half the amount calculated at the small $R_{\mathrm{MT}}$ value. We have noted in our previous study that such sensitivity to the size of projection space is amplified with a larger $U_{\mathrm{eff}}$ \cite{Park2024}. For example with $U_{\mathrm{eff}}$ $= 8$ eV (blue bars), the phase stability decreases more rapidly with increasing projection size and eventually reverses between $R_{\mathrm{MT}} =$ $2.1$ and $2.2$ $a_B$. On the other hand, using the projection-consistent $U_{\mathrm{eff}}$ values for rutile and anatase, anatase remains consistently more stable than rutile, with no clear trend across most of the range—particularly between $1.91$ to $2.30$ $a_B$ (black bars). 

\subsection{Properties of $\beta$-MnO$_2$ Calculated using Projection-consistent $U_{\mathrm{eff}}$ Scheme}

To examine whether the projection-consistent $U_{\mathrm{eff}}$ scheme also improves DFT+U calculations for systems with different $d$-electron occupancies and magnetic properties, we investigated the isostructural oxide $\beta$-MnO$_2$. The $U_{\mathrm{eff}}$ values for MnO$_2$  show the same characteristic trend with increasing $R_{\mathrm{MT}}$; they decrease from 5.6~eV at $R_{\mathrm{MT}} = 1.90\,a_B$ to 4.2~eV at $R_{\mathrm{MT}} = 2.20\,a_B$ (Fig.~\ref{fig:mno2_U}(a) and Table S4 in Supplementary Materials). This 25 \% reduction reflects the same dependence on the radial extent of the local orbitals as observed in TiO$_2$. Tompsett \textit{et al.} \cite{Tompsett2012} reported $U_{\mathrm{eff}} = 5.50$~eV using Wien2k, presumably at $R_{\mathrm{MT}} = 2.01\,a_B$ and employing a similar cDFT methodology. Their value is about $0.3~$ eV higher than ours ($5.2$ eV at $R_{\mathrm{MT}} = 2.00\,a_B$), though the limited details provided for their computational setup make this small discrepancy difficult to fully resolve. Nevertheless, despite these minor differences, the strong sensitivity of $U_{\mathrm{eff}}$ to the size of the projection space is clearly demonstrated.

\begin{figure}[h]
    \centering
    \includegraphics[width=0.7\textwidth, keepaspectratio]{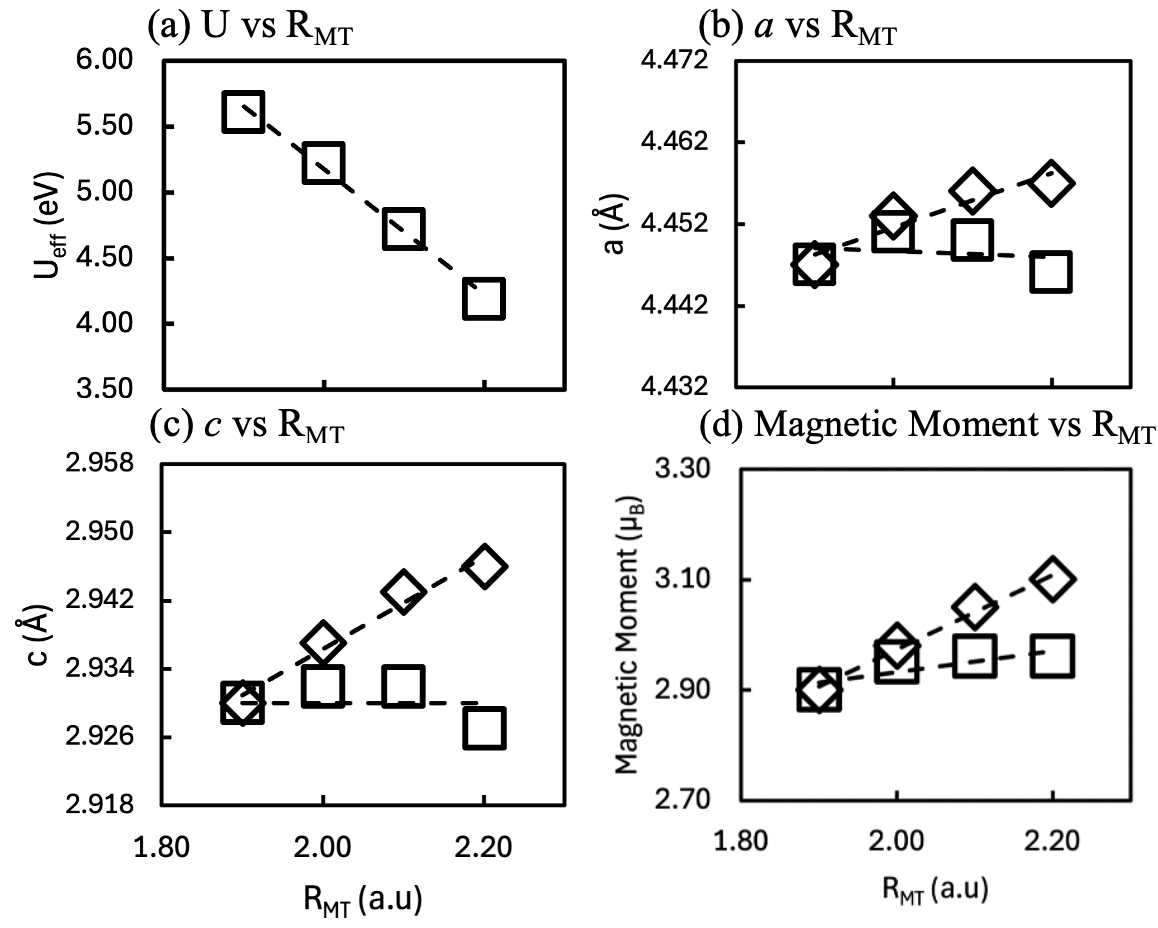}
    \caption{(a) \textit{ab initio} determined $U_{\mathrm{eff}}$ at selected Mn muffin-tin radii ($R_{\mathrm{MT}}$), (b) and (c) the calculated lattice parameters $a$ and $c$ (d) the magnetic moment as a function of Mn $R_{\mathrm{MT}}$ for $\beta$-MnO$_2$ in AFM ordering, respectively. In the panels (b) through (d), diamond markers correspond to the results using the fixed $U_{\mathrm{eff}}$ scheme, and square markers to those obtained under the projection-consistent $U_{\mathrm{eff}}$ scheme. Dashed lines are guides to the eye.
}

    \label{fig:mno2_U}
\end{figure}

Our results are consistent with the sensitivity of $U_{\mathrm{eff}}$ to projection space in $\beta$-MnO$_2$,previously hinted by Mahajan \textit{et al.} \cite{mahajan2021importance} in their study using pseudopotentials constructed from GBRV with PBEsol exchange-correlational functional. They reported that $U_{\mathrm{eff}}$ for $\beta$-MnO$_2$ varies by roughly 1.4--2.2 eV across different magnetic configurations when switching from non-orthogonalized atomic orbitals (NAO) to orthogonalized atomic orbitals (OAO). For example, $U_{\mathrm{eff}}$ values of 4.93 eV (FM) and 4.67 eV (A1-AFM) obtained using NAOs increased to 7.08 eV and 6.34 eV, respectively when OAOs were used. Because orbital orthogonalization--especially via L{\"o}wdin’s symmetric scheme--tends to contract the radial extent of the localized orbitals, this strong dependence of $U_{\mathrm{eff}}$ can be interpreted as another manifestation of projection-size-dependent Coulomb interaction.

The choice between fixed and projection-consistent $U_{\mathrm{eff}}$ schemes strongly influences structural and magnetic responses to changes in projection space. Using a fixed $U_{\mathrm{eff}} = 5.6$ eV at $R_{\mathrm{MT}} = 1.90$ $a_B$, structural optimization yields the lattice parameters of $a = 4.447 \, \angstrom$ and $c = 2.930 \, \angstrom$ for the AFM configuration (Fig.~\ref{fig:mno2_U}(b) and (c)). Increasing the projection size to $R_{\mathrm{MT}} =$ $2.20$ $a_B$ leads to larger lattice constants of $4.457 \, \angstrom$ and $2.946 \, \angstrom$, respectively (diamonds). In contrast, applying projection-consistent $U_{\mathrm{eff}}$ values at each $R_{\mathrm{MT}}$ keeps the lattice constants nearly unchanged, demonstrating the reduced sensitivity to the projection-space size when the projection-consistency is properly accounted for (squares). The magnetic properties show a similarly contrasting behavior. Under the fixed $U_{\mathrm{eff}}$ scheme, the Mn magnetic moment within the MT sphere increases from $2.90 \,\mu_B$ to $3.10 \,\mu_B$ as $R_{\mathrm{MT}}$ grows (Fig.~\ref{fig:mno2_U}(d)). The projection-consistent $U_{\mathrm{eff}}$ scheme clearly suppresses this trend, stabilizing the magnetic moment across MT radii. 

\begin{figure}[h]
    \centering
    \includegraphics[width=0.8\textwidth, keepaspectratio]{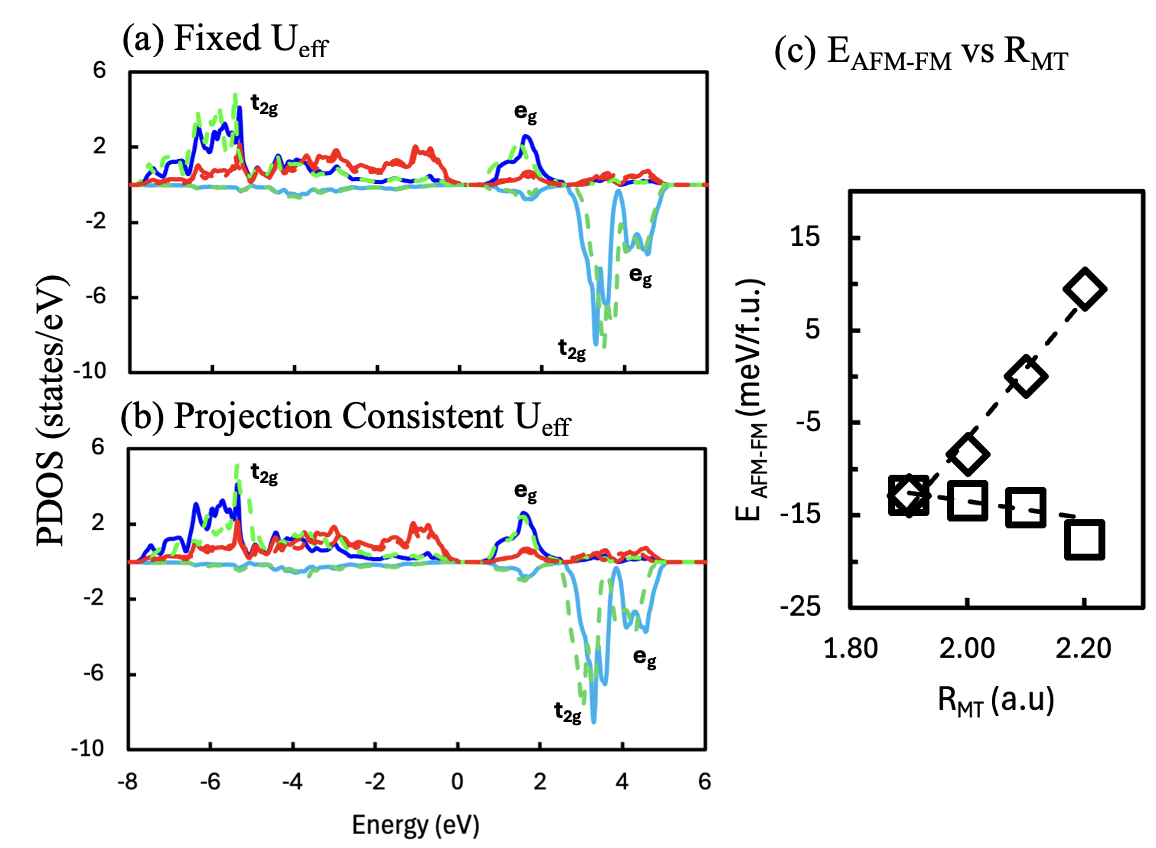}
    \caption{PDOS for Mn 3$d$ and O 2$p$ orbitals in $\beta$-MnO$_2$ using (a) the fixed and (b) projection-consistent $U_{\mathrm{eff}}$ schemes for two Mn $R_{\mathrm{MT}} =$ $1.90$ and $2.20$ $a_B$. For $R_{\mathrm{MT}} = $ $1.90$ $a_B$, PDOS are plotted in solid lines with Mn majority- and minority-spin states in dark and light blue, respectively as well as the O 2$p$ in red. For $R_{\mathrm{MT}} = $ $2.20$ $a_B$, PDOS are plotted in dashed lines with the same color scheme. (c) The total energy difference between AFM and FM configurations ($E_{AFM}- E_{FM}$) as a function of Mn $R_{\mathrm{MT}}$ calculated with the two schemes: using the fixed $U_{\mathrm{eff}}$ (diamonds) and projection-consistent $U_{\mathrm{eff}}$ values (squares). The dashed line are only to guide the eyes.}
    \label{fig:PDOS_MnO2_U_Rmt}
\end{figure}

The effects of the fixed and projection-consistent $U_{\mathrm{eff}}$ schemes on the spin-resolved Mn $d$ and O $p$ PDOS are shown in Figure~\ref{fig:PDOS_MnO2_U_Rmt}. Under both schemes using $R_{\mathrm{MT}} = 1.90$ $a_B$, the pronounced peaks between $-8$ and $-5$ eV correspond primarily localized $t_{2g}$ orbitals of majority-spin (solid blue line). The sharp feature near $-5.5$ eV is identified as the $d_{x^2 - y^2}$ orbital. This feature arises an octahedral distortion associated with the O-Mn-O bond angle in the equatorial plane to $\sim 80^{\circ}$ according to Tompsett \textit{et al}.\cite{Tompsett2012} The peaks around $2$ eV above the Fermi energy originate from the $e_g$ states of majority spin. The broad overlap between $e_g$ orbitals and O $p$ states (red) in the valence band region---more clearly visible with $R_{\mathrm{MT}} = 2.2$ $a_B$ (Fig. S1 in Supplementary Materials) indicates strong Mn–O covalency. Minority-spin $t_{2g}$ and $e_g$ orbitals lie above $E_F$ and remain largely unoccupied (light blue). 

Increasing the muffin-tin radius $R_{\mathrm{MT}} =$ $2.20$ $a_B$ while keeping $U_{\mathrm{eff}} = 5.6$ eV raises the minority-spin $t_{2g}$ levels by approximately $0.19$ eV (green dashed line), indicating reduced hybridization (Fig.~\ref{fig:PDOS_MnO2_U_Rmt} (a)). In contrast, under the projection-consistent scheme, these orbitals shift downward by about $0.26$ eV (Fig.~\ref{fig:PDOS_MnO2_U_Rmt} (b)). At the same time, the majority-spin $t_{2g}$ bands broaden toward higher energy, strengthening Mn-O overlap. Quantitative analysis of orbital charges and net spin occupations supports this interpretation (Table S5 in the Supplementary Materials).
The occupancy in minority-spin $t_{2g}$ increases by $32$ \% as $R_{\mathrm{MT}}$ grows from $1.90$ to $2.20$ $a_B$ under the projection-consistent $U_{\mathrm{eff}}$ scheme, whereas it increases only by $12$ \% when using the fixed $U_{\mathrm{eff}}$ scheme. These markedly different occupancies at $R_{\mathrm{MT}} = 2.20$ $a_B$ arise from the differing strengths of on-site Coulomb potentials; the fixed $U_{\mathrm{eff}}$ scheme overestimates this potential at larger projection size, suppressing orbital occupancy and thereby reducing hybridization. Ignoring the projection-size dependence of $U_{\mathrm{eff}}$ therefore leads to systematically weakened Mn-O hybridization at larger $R_{\mathrm{MT}}$.

The weakened hybridization between Mn $d$--O $p$ has previously been invoked to explain the AFM-to-FM transition that occurs as $U_{\mathrm{eff}}$ increases. For example, Crespo and Seriani \cite{crespo2013electronic} examined a range of AFM and FM configurations in a Hubbard-corrected DFT study of $\alpha$-MnO$_2$. They found that the FM ordering becomes energetically favored when $U_{\mathrm{eff}} \ge 2$~eV, whereas AFM ordering is correctly predicted for smaller values ($U_{\mathrm{eff}} \le 1.6$~eV). They attributed the stabilization of FM ordering to the weakening of AFM superexchange mediated by O $p$-orbitals, following a simplified argument consistent with the Goodenough--Kanamori--Anderson rules \cite{goodenough1955theory, goodenough1958interpretation, kanamori1959superexchange, anderson1950antiferromagnetism}. In particular, they proposed the reduced hybridization between Mn $d$ states and O $p_z$ orbitals---arising from $sp^2$-like mixing---diminishes AFM coupling and favors FM alignment.

An analogous AFM-to-FM transition is observed in our calculations as $R_{\mathrm{MT}}$ increases from $1.90$ to $2.20$ $a_B$ when using the fixed $U_{\mathrm{eff}}$ scheme (Fig.~\ref{fig:PDOS_MnO2_U_Rmt} (c)). At $R_{\mathrm{MT}} =$  $1.90$ $a_B$, AFM ordering is preferred by approximately $-13$ meV, but this preference steadily diminishes and ultimately reverses, yielding a $+9$ meV destabilization over FM ordering at $R_{\mathrm{MT}} = 2.20$ $a_B$ (diamonds). In contrast, calculations employing the projection-consistent $U_{\mathrm{eff}}$ scheme consistently predict an AFM ground state across the entire range of $R_{\mathrm{MT}}$values, with no transition to FM (squares). This behavior underscores the importance of using projection‑space‑dependent $U_{\mathrm{eff}}$ values to avoid artificial magnetic instabilities. 

There appears to be a clear equivalence between the effects of enlarging the projection space at a fixed $U_{\mathrm{eff}}$ and increasing $U_{\mathrm{eff}}$ at a fixed projection size. For $\beta$-MnO$_2$, Wang \textit{et al.} \cite{wang2016local} reported that FM ordering becomes favored when the $U_{\mathrm{eff}}$ exceeds about $5$ eV, provided a sufficiently large muffin-tin radius---mirroring our observation of an AFM-to-FM transition at larger $R_{\mathrm{MT}}$ when a fixed $U_{\mathrm{eff}}$ is used. Franchini \textit{et al.}~\cite{franchini2007ground} likewise found that increasing $U_{\mathrm{eff}}$ from 3 to 6~eV leads to an expansion of the unit-cell volume and an increase in magnetic moment from 2.93~$\mu_B$ to 3.34~$\mu_B$. These trends closely match the lattice expansion and enhanced magnetic moments we observe as  $R_{\mathrm{MT}}$ increases (Fig. \ref{fig:mno2_U} (b) through (d)). A similar correspondence is seen in TiO$_2$, where the increase in lattice parameters with larger $R_{\mathrm{MT}}$ under the fixed $U_{\mathrm{eff}}$ scheme (Fig.~\ref{fig:latticevsRmt}) parallels the behavior reported when $U_{\mathrm{eff}}$ is varied at constant projection size.~\cite{dompablo2011dft+,Park2024} Likewise the dependence of crystal-field splitting on $R_{\mathrm{MT}}$ (Fig.~\ref{fig:PDOS_rutile_U_cal}) closely resembles its sensitivity to $U_{\mathrm{eff}}$ at fixed projection size.~\cite{dompablo2011dft+,Park2024,zhao2018does} 

\section{Discussion}

\begin{figure}[h]
    \centering
    \includegraphics[width=0.6\textwidth, keepaspectratio]{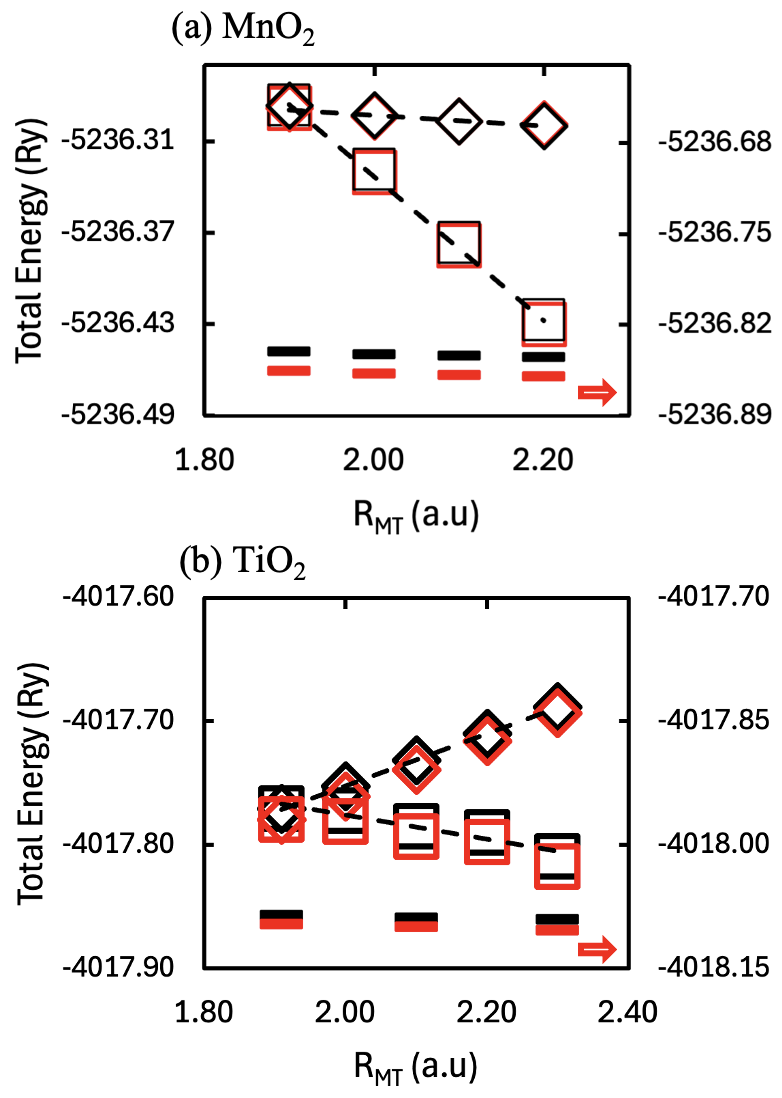}
    \caption {Comparison of the total energy as a function of $R_{\mathrm{MT}}$ between the fixed (diamonds) and projection-consistent \( U_{\mathrm{eff}} \) (squares) schemes in (a) the AFM (red) and FM (black) configurations for MnO$_2$ and (b) for TiO$_2$. Under the fixed \( U_{\mathrm{eff}} \) scheme, the values of $5.6$ and $4.5$ eV are used for MnO$_2$ and TiO$_2$ calculations, respectively. Also shown are the total energy calculations without the Hubbard correction (bars).}
    \label{fig:TotEvsRmt}
\end{figure}

The use of internally consistent $U_{\mathrm{eff}}$ values can evidently alleviate the sensitivity of not only electronic properties, but also optimized lattice parameters and relative phase stability, to the size of the projection space. In order to gain a broader understanding into the dependence of the DFT+U calculations on the projection size, we compare the total energy calculated by the fixed and projection-consistent $U_{\mathrm{eff}}$ schemes. Figure \ref{fig:TotEvsRmt} (a) shows the total energies of MnO$_2$ calculated at various $R_{\mathrm{MT}}$ values using the fixed $U_{\mathrm{eff}} =$ $5.6$ eV (diamonds) and the projection-consistent values (squares), for both AFM (red) and FM (black) configurations. At $R_{\mathrm{MT}} =$ $1.90$ $a_B$, the AFM configuration is slightly lower in energy than the FM configuration. Under the fixed $U_{\mathrm{eff}}$ scheme, total energies of both AFM and FM states decrease slightly as $R_{\mathrm{MT}}$ increases; however, the FM energy decreases more rapidly, ultimately reversing the relative stability of the two magnetic states.

In contrast, when projection-consistent $U_{\mathrm{eff}}$ values are used, the total energies of the AFM and FM configurations decrease at nearly the same rate with increasing $R_{\mathrm{MT}}$. As a result, the energy difference between the two magnetic states remains essentially constant across the entire projection-space range, and the AFM ground state is preserved. Notably, the overall decrease in total energy is more pronounced under the projection-consistent $U_{\mathrm{eff}}$ scheme than under the fixed scheme. For both rutile and anatase TiO$_2$ (Fig. \ref{fig:TotEvsRmt} (b)), the behavior is reversed: the total energy obtained with the projection-consistent $U_{\mathrm{eff}}$ decreases only moderately as $R_{\mathrm{MT}}$ increases (squares), whereas the total energy computed using the fixed $U_{\mathrm{eff}}$ scheme rises sharply with increasing projection size (diamonds). Despite the reversed trends in total energy, it is the projection‑consistent $U_{\mathrm{eff}}$ scheme that yields nearly constant energy differences as $R_{\mathrm{MT}}$ increases.

For both MnO$_2$ and TiO$_2$, calculations performed with the PBE functional alone show no significant dependence of the total energy on $R_{\mathrm{MT}}$ (bars in Fig. \ref{fig:TotEvsRmt} (a) \& (b)). This indicates that the strong dependence observed in the DFT+U calculations arises primarily from the Hubbard correction term. The contrasting responses of the total energy as a function of $R_{\mathrm{MT}}$ can be understood by examining the Hubbard energy contribution $E[n(\mathbf{r})]$ in Eq. \ref{eq:Hubbard}. Using the Hellmann--Feynman theorem, following the approach of Solovyev and Dederichs\cite{Solovyev1994}, the dependence of the interaction energy on the projection size $R$ through both $n$ and $U_{\mathrm{eff}}$ can be expressed as
\begin{equation}
\frac{\partial E}{\partial R} = U_{\mathrm{eff}} \, (\frac{1}{2}-n) \frac{\partial n}{\partial R} + \frac{n(1-n)}{2} \frac{\partial U_{\mathrm{eff}}}{\partial R}
\label{eq:dEdR}
\end{equation}

In the fixed $U_{\mathrm{eff}}$ scheme, the second term is effectively neglected by assuming $\frac{\partial U_{\mathrm{eff}}}{\partial R} = 0$. Because $\frac{\partial n}{\partial R}$ is generally positive, the sign of the first term is determined by the occupancy of the $d$-orbitals. In MnO$_2$, the combined spin-up and spin-down occupancy $n = n_\uparrow + n_\downarrow$ exceeds one half (Table~S6), resulting in a negative first term with $(\frac{1}{2}-n) < 0$, and thus a decreasing total energy with increasing $R_{\mathrm{MT}}$ (diamonds in Fig. \ref{fig:TotEvsRmt} (a)). In TiO$_2$, the $d$-orbitals are less than half-filled, so $(\frac{1}{2}-n) > 0$ (Table~S3). This leads to a monotonically increasing total energy with increasing projection size, as observed in Fig.~\ref{fig:TotEvsRmt}(b) (diamonds). 

Under the projection-consistent $U_{\mathrm{eff}}$ scheme, the second term of Eq.~\ref{eq:dEdR} is no longer neglected. This term is generally negative, since $\frac{\partial U_{\mathrm{eff}}}{\partial R} < 0$, reflecting the weakening of the Coulomb repulsion with increasing projection size, as shown in Figs. \ref{fig:mno2_U}(a), \ref{fig:UvsRmt}(a), and \ref{fig:UvsRmt}(b) for MnO$_2$ and TiO$_2$, respectively. For MnO$_2$, this negative second term reinforces the already decreasing trend of the total energy (squares in Fig. \ref{fig:TotEvsRmt} (a)). In contrast, for TiO$_2$, the negative second term competes against the positive contribution from the first term of Eq.~\ref{eq:dEdR}, resulting in a more moderate decrease in total energy (squares in Fig. \ref{fig:TotEvsRmt} (b)).

The contrasting total-energy trends with increasing muffin‑tin radius demonstrate that the projection dependence of the Hubbard correction can drive qualitatively different energetic responses. While these results clearly establish the importance of projection consistency for total energies, they do not by themselves identify the microscopic origin of this behavior. To elucidate the physical mechanism underlying the systematic reduction of $U_{\mathrm{eff}}$ with increasing projection size—and its role in stabilizing projection‑independent energetics—we now examine how the spatial characteristics of the localized $d$ orbitals evolve as the size of projection space is varied.

\begin{figure}[h]
    \centering
    \includegraphics[width=0.6\textwidth, keepaspectratio]{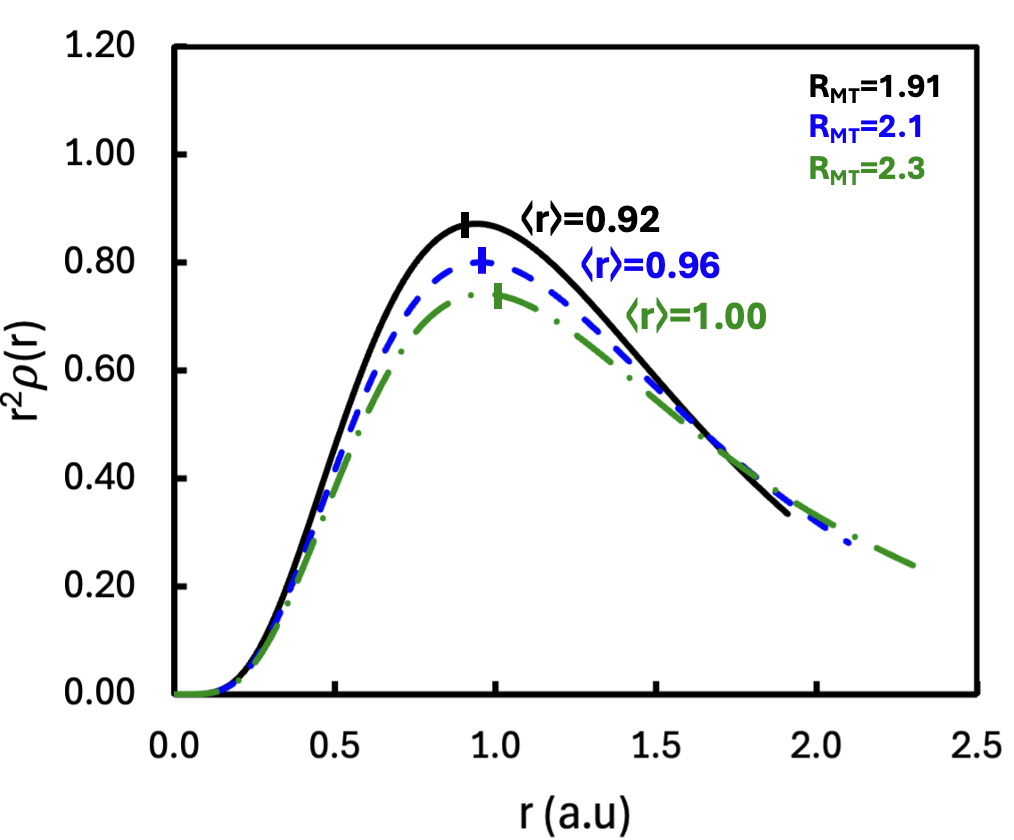}
    \caption{Radial probability distribution function of Ti $d$ orbitals in rutile TiO$_2$ calculated for three different muffin-tin radii, using the PBE exchange-correlation functional.}
    \label{fig:Radial_Prob}
\end{figure}

Figure \ref{fig:Radial_Prob} shows the radial probability distribution $r^2\rho(r)$ of the Ti $d$ orbitals in rutile TiO$_2$, calculated using the PBE functional, for three different values of $R_{\mathrm{MT}}$. As $R_{\mathrm{MT}}$ increases, the shape of the Ti $d$ orbitals evolves noticeably, with a more pronounced tail extending outward for larger projection spaces. The mean radius of the $d$ orbitals $\langle r \rangle = \int_0^{R_{MT}} r^2 dr \rho(r) r$ provides a useful quantitative measure of this spatial extent. As $R_{\mathrm{MT}}$ increases from $ 1.91$ to $2.30$ $a_B$, the $\langle r \rangle$ increases from $0.92$ $a_B$ to $1.00$ $a_B$, respectively, indicating a clear spatial dilation of the localized orbitals.    

This increase in spatial extent directly affects the computed on-site Coulomb interaction $U_{\mathrm{eff}}$. It is important to recall that $U_{\mathrm{eff}}$ represents the screened energy cost of transferring a localized electron from one site to another.\cite{herring1966magnetism,Anisimov1991} This energy cost includes screening by other valence electrons such as the 4$s$ and 4$p$ electrons in 3$d$ transition metals. Such screening reduces $U_{\mathrm{eff}}$ in the solid much smaller than the unscreened Slater integral of the corresponding atom.\cite{Madsen2005} Solovyev and Dederichs also highlighted the importance of orbital relaxation in understanding on-site Coulomb interaction $U_{\mathrm{eff}}$.\cite{Solovyev1994} They showed that the $U_{\mathrm{eff}}$ of transition metal impurities in Rb varies significantly with electronic configuration: for $3d$ impurities, the value is about $30$ $\%$ larger in the divalent configuration than in the monovalent one. They attributed this dependence to changes in relaxation and screening as the localized orbitals become more or less contracted---an effect that mirrors the sensitivity of $U_{\mathrm{eff}}$ on the atomic number for a fixed valence configuration. 

The observed dependence of $U_{\mathrm{eff}}$ on the $R_{\mathrm{MT}}$ values as well as the basis sets reflects this relaxation and screening mechanism. As the projection space expands, the mean separation between the localized electrons increases accordingly. The \textit{ab initio} $U_{\mathrm{eff}}$ value calculated for increasing projection size therefore decreases to exhibit the reduced on-site Coulomb repulsion. Consequently, $U_{\mathrm{eff}}$ should be recalculated or re-scaled for each chosen projection space, not based on an empirical and pragmatic observation but on the physical reasoning that the Hubbard interaction term correctly incorporates the internally consistent Coulomb interaction within the DFT+U framework. 

\section{Conclusions}
The sensitivity of DFT+U predictions to the choice of projection space has long been recognized as a practical limitation, yet its physical origin and energetic consequences have remained incompletely characterized. In this work, we have shown that the effective on-site Coulomb interaction $U_{\mathrm{eff}}$, determined \textit{ab initio}, decreases systematically as the size of the local projection space is increased. This reduction can reach approximately 30\% in both strongly covalent systems with nearly empty $d$ shells (TiO$_2$) and partially filled $d$-electron systems with competing magnetic interactions ($\beta$-MnO$_2$).

We demonstrated that applying a fixed $U_{\mathrm{eff}}$ calibrated at one projection size while varying the projection space leads to physically inconsistent Hubbard corrections, producing artificial total-energy trends, spurious magnetic phase transitions, and projection-dependent phase stabilities. These pathologies arise from the combined projection-space dependence of orbital occupancies and the screened Coulomb interaction. By contrast, determining $U_{\mathrm{eff}}$ in an internally consistent manner for each projection space yields energetically robust predictions that are largely insensitive to the size of the projection.

Analysis of the orbital character and total-energy response shows that the reduction of $U_{\mathrm{eff}}$ with increasing projection size is a direct consequence of orbital relaxation and enhanced electronic screening associated with the spatial expansion of the localized orbitals. This mechanism is consistent with early theoretical analyses of screened Coulomb interactions and provides a unified physical explanation for similar trends observed across different computational frameworks.

More broadly, these results clarify the apparent non-transferability of reported $U_{\mathrm{eff}}$ values in the literature: differences in $U_{\mathrm{eff}}$ largely reflect differences in the effective projection space sampled by each method. The projection-consistent framework introduced here provides a physically transparent and computationally tractable strategy for reconciling these values and for obtaining reliable energetic predictions in DFT+U calculations. We anticipate that this approach will be particularly valuable in studies of phase stability, magnetism, and other total-energy-derived properties, as well as in high-throughput and cross-code applications where control of projection-space artifacts is essential.

\section{Acknowledgements}
The authors are grateful to P. Blaha and D. O'Regan for discussion on various aspects of the manuscript. M.R. is thankful for the travel support from the Department of Physics and Astronomy and the School of Graduate Studies at Baylor University. The authors acknowledge C. Bell at the High Performance Computing Center of Baylor University for technical support.

\section{}
\bibliographystyle{unsrt}
\bibliography{Abating_the_sensitivity}

\end{document}